\documentclass[12pt]{article}
\pdfoutput=1
\usepackage{ifpdf}
\usepackage[a4paper,top=3.3cm,width=17cm, bottom=3.3cm, left=2cm]{geometry}

\usepackage{fancyhdr}
\usepackage[utf8x]{inputenc}
\usepackage{graphicx}
\usepackage[font=small,labelfont=bf]{caption}
\usepackage{microtype}

\usepackage{authblk}

\title{{\bf Statistics of thermalization in Bjorken Flow}}
\author[1]{Jakub Jankowski\footnote{Email: jakubj@th.if.uj.edu.pl}}
\author[2]{Grzegorz Plewa\footnote{Email: g.plewa@ncbj.gov.pl}}
\author[2,3]{Micha\l\ Spali\'nski\footnote{Email: mspal@fuw.edu.pl}}
\affil[1]{Institute of Physics, Jagiellonian University, ul. {\L}ojasiewicza 11, 30-348 Krak\'ow, Poland}
\affil[2]{National Center for Nuclear Research, ul. Ho\.za 69, 00-681 Warsaw, Poland}
\affil[3]{Physics Department, University of Bia{\l}ystok, ul. Lipowa 41, 15-424 Bia{\l}ystok, Poland}
\date{}

\newcommand{\p}{\partial}

\newcommand{\bea}{\begin{eqnarray}}
\newcommand{\beal}[1]{\begin{eqnarray}\label{#1}}
\newcommand{\eea}{\end{eqnarray}} 
\newcommand{\be}{\begin{equation}} 
\newcommand{\bel}[1]{\begin{equation}\label{#1}}
\newcommand{\ee}{\end{equation}} 
\newcommand{\rf}[1]{(\ref{#1})}
\newcommand{\f}[2]{\frac{#1}{#2}}
\newcommand{\nn}{\nonumber}
\newcommand{\bit}{\begin{itemize}}
\newcommand{\eit}{\end{itemize}}
\newcommand{\ben}{\begin{enumerate}}
\newcommand{\een}{\end{enumerate}}

\newcommand{\sym}{${\mathcal N}=4$ SYM}

\newcommand{\bars}{S}
\newcommand{\barsi}{S_i}
\newcommand{\barsf}{S_f}
\newcommand{\ent}{{\cal S}}
\newcommand{\efft}{T}
\newcommand{\effti}{T_i}

\def\t{\tilde}

\def\half{\frac{1}{2}}

\def\alp{\leavevmode\ifmmode {\alpha^\prime} \else ${\alpha^\prime}$ \fi}

\def\ta{\t{a}}
\def\tb{\t{b}}
\def\tc{\t{c}}
\def\taz{\t{a}^{(0)}}
\def\tbz{\t{b}^{(0)}}
\def\tcz{\t{c}^{(0)}}

\begin{document}

\maketitle

\thispagestyle{empty}

\abstract{The apparent early thermalization of quark-gluon plasma produced at
  RHIC and LHC has motivated a number of studies of strongly coupled
  ${\mathcal N}=4$ supersymmetric Yang-Mills theory using the AdS/CFT
  correspondence. Here we present the results of numerical simulations of
  Bjorken flow aimed at establishing typical features of the dynamics. This is
  done by evolving a large number of far from equilibrium initial states well
  into the hydrodynamic regime.  The results strongly suggest that early
  thermalization is generic in this theory, taking place at proper times
  just over $0.6$ in units of inverse local temperature at that time. We also find that the
  scale which determines the rate of hydrodynamic cooling is linearly
  correlated with the entropy of initial states defined by the area of the
  apparent horizon in the dual geometry. Our results also suggest that
  entropy production during the hydrodynamic stage of evolution is not 
  negligible despite the low value of $\eta/s$. 
  }

\newpage

\section{Introduction}

Experimental studies of quark-gluon plasma at RHIC and LHC have led to a
number of theoretical challenges. The successful phenomenological description
of what is observed rests on the assumption that the system thermalizes on a
time scale of order less than inverse local temperature. By thermalization we
mean here that a hydrodynamic description becomes valid
\cite{Teaney:2000cw,Kolb:2003dz}. It has recently been appreciated that at
least in some model cases this happens while the system is still very
anisotropic \cite{Chesler:2009cy,Heller:2011ju}, so instead of using the term
thermalization one sometimes speaks of ``hydrodynamization''.

It is known that in a strongly interacting system the approach to 
hydrodynamics {\em can} be very rapid \cite{Heller:2011ju,Heller:201e2j} 
-- as rapid as seen in heavy ion collisions. One of the
purposes of the present work was to see how generic this behaviour is.  This
type of question was recently considered in the context of isotropization
\cite{Heller:2013oxa}, where no hydrodynamic modes are present. Here we
consider a case where we can see the onset of hydrodynamic behaviour (as in
\cite{Heller:2011ju,Heller:201e2j}). We consider a large number of far from
equilibrium initial states (which mimic the situation right after heavy nuclei
collide) and simulate the subsequent evolution until hydrodynamic behaviour
sets in. This way we can see, in particular, the distribution of
thermalization times.

While the theory underlying the phenomena studied at RHIC and LHC is firmly
believed to be textbook QCD, it has so far not been possible to understand the
observations from first principles. Since the post-collision plasma state
appears to be strongly coupled, much effort has been devoted to consideration
of similar physical issues in the context of strongly coupled
\sym\ \cite{CasalderreySolana:2011us}, which has a tractable holographic
representation in terms of AdS gravity \cite{Maldacena:1997re,Witten:1998qj}.
Even then it has not been possible to describe the entire process
analytically, making it necessary to resort to suitable numerical techniques.

Numerical studies of the dynamics of Yang-Mills plasma based on the AdS/CFT
correspondence have been an area of significant activity in recent years (for
a review and references see \cite{Chesler:2013lia}). Holography represents
strongly interacting Yang-Mills plasma (in $d=3+1$ dimensions) by a spacetime
geometry in one dimension higher. Initial states of the plasma are represented
as initial geometries which are then evolved numerically using Einstein's
equations.  These computations are in general rather daunting, so much effort
has been devoted to situations where symmetries reduce the complexity to a
manageable level. A physically important case is that of Bjorken flow
\cite{Bjorken:1982qr}. Despite considerable simplification of the dynamics,
this example is rich enough to be of phenomenological significance, even
though its usefulness is limited to the central rapidity region.

Studies of Bjorken flow in the context of AdS/CFT were initiated in
\cite{Janik:2005zt}, which showed analytically that hydrodynamic behaviour
appears at late times. The dynamics at early times was considered soon after
\cite{Beuf:2009cx}, but it quickly became clear that a numerical approach to
the problem was required. The first numerical studies were pioneered by
Chesler and Yaffe \cite{Chesler:2009cy}, but rather than setting general
initial conditions (corresponding to non-equilibrium states of the dual gauge
theory plasma) in the bulk the authors considered perturbing the system at the
boundary.

Formulating consistent initial conditions requires solving the constraints
contained in Einstein equations. This rather nontrivial problem was considered
in \cite{Beuf:2009cx}, where 29 solutions were found.  These initial
conditions were the starting point for the numerical studies
\cite{Heller:2011ju,Heller:201e2j}, where a number of important observations
were made. Some of the results were however limited by the fact that only a
small number of initial conditions were available.

The present work follows \cite{Chesler:2008hg,Chesler:2009cy} in using
Eddington-Finkelstein coordinates. The parameterization of the metric used
here is different however, because the original approach of
\cite{Chesler:2009cy} made it difficult to set the initial conditions at very
early proper time. The approach described here overcomes this difficulty. The
form of the field equations makes it possible to effectively solve the
constraints. Initial conditions can be set by choosing a single independent
metric coefficient function which satisfies some mild conditions and the
remaining metric coefficients on the initial time slice are then determined by
numerically integrating ordinary differential equations.

Given a well posed initial value problem it is possible to solve the field
equations numerically. The complications of AdS asymptotics can be dealt with
following the approach of \cite{Chesler:2009cy} (see also
\cite{Murata:2010dx}). Once the spacetime geometry is determined (up to some
final time) the AdS/CFT prescription provides an explicit recipe to determine
the expectation value of the energy-momentum tensor of the dual 4-dimensional
gauge theory.

At sufficiently late times the system is governed by the equations of
hydrodynamics -- this was shown analytically in \cite{Janik:2005zt} and
numerically in \cite{Heller:2011ju,Heller:201e2j}. The determination of the
``typical'' time when hydrodynamic behaviour sets in is one of the main tasks
we address. Heller et al. \cite{Heller:2011ju,Heller:201e2j} analysed 29
initial conditions which indicated the hydrodynamics sets in quite early and
noted that the pressure anisotropy at that time is still significant (this was
also pointed out in a similar context by \cite{Chesler:2009cy}). The approach
adopted in this paper makes it possible to generate consistent initial
conditions at will, so we are able to look at a large number of
solutions. This makes it possible to claim a much more firm estimation. In
what follows we describe the distribution of thermalization times for a sample
of over 600 initial conditions.

An important role in the interpretation of our simulations is played by
entropy.  After thermalization the hydrodynamic notion of entropy
\cite{Bhattacharyya:2008xc,Booth:2009ct,Booth:2010kr} is valid, but also at
earlier times the apparent horizon\footnote{The boundary-covariant apparent
  horizon consistent with conformal symmetry is unique in fluid-gravity
  duality\cite{Booth:2011qy}.} area (mapped to the boundary) gives rises to an
operational notion of entropy, which asymptotes to
hydrodynamic entropy at late times. 

The organization of the paper is as follows. Section~\ref{sec.bif} 
recalls the implications of boost invariance and the hydrodynamics of Bjorken
flow. In section~\ref{sec.gravity} we present the dual, gravitational
description and some details of the numerics. Initial states for the
simulation are discussed in section~\ref{sec.initial}. The results for
thermalization are presented in section~\ref{sec.thermal} and entropy
production is the subject of section~\ref{sec.entropy}.


\section{Boost invariant flow}
\label{sec.bif}

An idealized, but very useful description of a heavy ion collision was
formulated by Bjorken \cite{Bjorken:1982qr}.  The colliding nuclei are
regarded as infinite sheets of mater extending in the plane transverse to the
collision axis. The collision energy is also infinitely high, and the physics
of the system is assumed to be determined only by the proper time evolution of
the energy density. In the case of central, ultrarelativistic collisions this
is a useful approximation for the description of observables in central
rapidity region.

Boost invariant states in four spacetime dimensions can most
simply be characterized by saying that in proper time-rapidity coordinates 
\be
t = \tau \cosh y \ , \quad
z = \tau \sinh y \ ,
\ee
all observables are independent of the rapidity $y$. Furthermore, we assume
independence of the transverse coordinates. 

In particular, in a conformal theory the expectation value of the energy
momentum tensor in a 
boost-invariant state takes the form \cite{Janik:2005zt}
\bel{Tab}
T_{\mu\nu}= \mathrm{diag}(\epsilon, p_L, p_T, p_T) \ ,
\ee
where
\bel{Pr}
p_L = - \epsilon - \tau\dot{\epsilon}\ , \quad 
p_T =  \epsilon + \half \tau\dot{\epsilon} \ ,
\ee
so it is determined completely by the energy density
$\epsilon(\tau)$. Furthermore $\epsilon\sim T^4$, and for 
\sym\ \cite{Witten:1998zw} in equilibrium 
\bel{efftemp}
\epsilon =\frac{3}{8} \pi^{2} N_c^2 T^4 \ .
\ee
It is
very convenient to introduce the notion of \emph{effective temperature}, 
which satisfies eq.~\rf{efftemp} in any state; thus it is the temperature of an
equilibrium state with the same energy density \cite{Heller:2007qt}. In what
follows we will denote the effective temperature by $T$, since this introduces
no ambiguity\footnote{Note that it may not be possible to do this for more
  general flows \cite{Arnold:2014jva}.}.  

In a hydrodynamic state, by definition, the energy momentum tensor can be
expressed in terms of the local fluid velocity $u$ and temperature
$T$. Specifically, it can be written in the form 
\be
\langle  T^{\mu\nu} \rangle = \epsilon u^\mu u^\nu + p ( u^\mu u^\nu +
\eta^{\mu\nu}) + \dots 
\ee
where 
the ellipsis denotes terms involving gradients of the hydrodynamic
variables. The energy density $\epsilon$ and pressure $p$ are understood to be
expressed in terms of the temperature using the equations of state. For a
conformal field theory one has $\epsilon=3 p\sim T^4$.

A boost invariant configuration  has $(u^\mu)=(-1,0,0,0)$, and all the physics is
encoded in the dependence of the energy density (or, equivalently, the
temperature) on $\tau$. Analytic calculations 
have been performed up to third order in the gradient expansion
\cite{Janik:2006ft,Heller:2007qt,Booth:2009ct}, the result being 
\beal{thydro}
T(\tau) &=& \frac{\Lambda}{\left( \Lambda \tau \right)^{1/3}} \Big\{ 1 -
\frac{1}{6 \pi \left( \Lambda \tau \right)^{2/3}} +  
\frac{-1 + \log{2}}{36 \pi^{2} \left( \Lambda \tau \right)^{4/3}} +\nn\\
&+& \frac{-21 + 2\pi^{2} + 51 \log{2} - 24 \log^{2}{2}}{1944 \pi^{3} \left(\Lambda \tau\right)^{2}}
\Big\}~.
\eea
The scale $\Lambda$ appearing in \rf{thydro} depends on the initial
conditions. Indeed, it is the only trace of the initial conditions to be found
in the late time behaviour of the system. 
It is expected that starting from any initial state the system should 
evolve in such a way that at late times it will 
be described by \rf{thydro} for some value of the constant $\Lambda$. As we shall
see, our simulations suggest that for ``low entropy'' initial states $\Lambda$
is proportional to the apparent horizon entropy of the initial state.

To determine whether 
that the system has thermalized (in the sense of reaching hydrodynamics) it is
very convenient to define the function \cite{Heller:2011ju} 
\bel{fdef}
f(w) = \f{\tau}{w} \f{dw}{d\tau} \ ,
\ee 
where $w=\tau\efft(\tau)$ is the proper time in units of inverse
effective temperature. In a conformal theory the behaviour of this 
function at large $w$ is independent of
$\Lambda$ for dimensional reasons. By computing
it at late times (using \rf{thydro}) one finds
\bel{fhydro}
f_H(w) = \f{2}{3}+ \f{1}{9\pi w} +
\f{1-\log 2}{27\pi^2 w^2}+
\f{15-2\pi^2-45\log 2+24 \log^2 2}{972 \pi^3 w^3} +\ldots \ .
\ee
A given solution, $\epsilon(\tau)$, starts out far from equilibrium, but the
damping of non-hydrodynamic modes ensures that after a sufficiently long time
only hydrodynamic modes remain. Given the result of a numerical simulation for
$T(\tau)$ one 
can check whether the 
hydrodynamic regime has been reached by comparing $f(w)$ calculated from 
eq.~\rf{fdef} with the hydrodynamic form given by eq.~\rf{fhydro}. 

An important role in the following is played by entropy of the expanding
plasma. The thermodynamic (equilibrium) entropy density is given by 
\bel{eqed}
s = \frac{1}{2} \pi^{2} N_c^2 T^3 \ .
\ee
In a boost invariant setting it is useful to speak of entropy per unit
rapidity and transverse area, given by 
\bel{entperura}
\f{d\ent}{dy dx_\perp^2} = \tau s \ ,
\ee
where the factor of $\tau$ reflects the expanding volume. 
In our holographic calculation the entropy per unit rapidity and transverse
area can be computed from the apparent horizon
area (due to the Bekenstein-Hawking relation), as
discussed in 
section~\ref{sec.entropy}. 

As discussed in
\cite{Heller:201e2j} it is convenient to use the dimensionless ratio  
\bel{entrat}
\bars = \f{1}{\frac{1}{2} \pi^2 N_{c}^2 \cdot  \big(\effti\big)^2} \Bigg(
\f{d\ent}{dy dx_\perp^2} \Bigg) \ ,
\ee
where $\effti$ is the effective temperature at the initial time. 
At late times, the ratio \rf{entrat} approaches the limiting value 
\bel{ethermo}
\label{entfinal}
\barsf = \Lambda^{2}  \big(\effti\big)^{-2} \ .
\ee
Henceforth when speaking of entropy, we will mean the quantity given in
eq.~\rf{entrat}.


\section{The gravity dual}
\label{sec.gravity}

The standard AdS/CFT prescription for calculating the energy-momentum
tensor expectation value in a given state is to find the gravity solution
corresponding to that state and then extract the expectation value from the
near-boundary asymptotics. 

The object of this paper is to simulate the approach to equilibrium starting
from generic, non-equilibrium initial states of Yang-Mills
plasma at $\tau=0$. In the holographic approach pursued here this is
implemented by choosing an initial geometry at random and
evolving it using Einstein's equations. 

The choice of initial geometry must
satisfy the constraint equations, whose form depends on the choice of
coordinates and parameterization of the metric. The choices adopted here draw on the
experience gained from references \cite{Chesler:2009cy,Heller:201e2j} and
make it possible to 
effectively solve the constraints at $\tau=0$, so that an arbitrary number of
consistent initial geometries can be generated and evolved
\cite{HellerSpalinski}. As in the work of Chesler and Yaffe
\cite{Chesler:2009cy,Chesler:2013lia} (as 
well as the papers on fluid-gravity duality \cite{Bhattacharyya:2008jc}), we
adopt Eddington-Finkelstein 
coordinates. The metric is parameterized as  
\bel{mansatz}
ds^2 = \frac{1}{z^2} \left(- 2 dz dt - a(t,z) dt^2 + b(t,z) dy^2 + c(t,z)
dx_\perp^2 \right)\ .
\ee
This parameterization is different from that used in \cite{Chesler:2009cy},
which is inconvenient if one wishes to set initial conditions at $t=0$. With the
parameterization \rf{mansatz} it is 
possible to set initial conditions at any $t$. Note that on the conformal
boundary this time
coordinate coincides with the proper time coordinate $\tau$.

Assuming the metric Ansatz given in eq.~\rf{mansatz}, the Einstein equations
with negative 
cosmological constant 
\bel{einstein}
R_{ab} + 4 g_{ab}=0
\ee
reduce to a set of 5 nonlinear partial differential equations for the metric
coefficients $a, b, c$. Three 
of these equations can be regarded as evolution equations. Of the remaining
two, one is a constraint on initial data (and will be discussed in detail in
section~\ref{sec.initial}), and the other is redundant in the
sense that it is automatically satisfied by any consistent initial data
evolved using the evolution equations. 

Due to the presence of a negative cosmological constant the solutions describe
locally asymptotically AdS geometries. For the
purpose of numerical computations it is very convenient to isolate the
asymptotic AdS behaviour by defining
\beal{regs}
a(z,t) &=& 1 + z^3\ \ta(z,t)\nn\\
b(z,t) &=& (t+z)^2 + z^3\ \tb(z,t)\nn\\
c(z,t) &=& 1 + z^3\ \tc(z,t) \ .
\eea
The evolution equations expressed in terms of $\ta, \tb, \tc$ are amenable to
numerical computations at $t\neq 0$, but at $t=0$ one has to set 
\beal{regsz}
a(z,t) &=& 1 + z^3\ \taz(z,t)\nn\\
b(z,t) &=& (t+z)^2 + z^5\ \tbz(z,t)\nn\\
c(z,t) &=& 1 + z^3\ \tcz(z,t) 
\eea
in order to obtain differential equations which can be 
used to make the initial time step at $t=0$. 

With the definitions \rf{regs}-\rf{regsz} the evolution equations assume a 
rather special form: even though they are nonlinear, they are linear in the
time derivatives\footnote{This is of course a consequence of using the
  Eddington-Finkelstein gauge.} . Specifically, letting $u$ denote the 4-component vector
$(\p_t\tb, \p_t\tc, \p_z\ta,\ta)$ 
one finds that the 4 independent Einstein equations can be written as 
\beal{linsys}
\p_z u = A u + f \ ,
\eea
where $A$ is a 4x4 matrix and $f$ is a 4-component vector expressed in terms
of $\ta,\tb,\tc$ and their $z$ 
derivatives. 
Solving this linear system of first order ordinary differential equations
by integration over $z$ one can 
obtain the time derivatives $\p_t\tb$, $\p_t \tc$ and $\ta$ itself. These can
then be used to evolve
the metric. The explicit form of $A$ and $f$ is unilluminating, but
straightforward to obtain. As mentioned earlier, for the time step at $t=0$
one needs to use \rf{regsz} instead of \rf{regs}.

The integration over $z$ can conveniently be carried out using the spectral
representation \cite{Grandclement:2007sb}, which we use in the form of the 
collocation method with Chebyshev polynomials. For regular configurations,
such as those considered here, it gives exponential convergence, allowing for
modest grid sizes. 

To solve eq.~\rf{linsys} one must impose the correct boundary
conditions. These can be 
determined by solving the field equations in a near boundary expansion,
i.e. an expansion around $z=0$. The result of this standard procedure is 
\beal{expan}
a(t,z) &=& 1 + a_4(t)   z^4 + 
\ldots\nn \\
b(t,z) &=& t^2 + 2 t z + z^2 + z^4\ t^2 \left(a_4(t) + \frac{3}{4} t a_4'(t)
\right) +
\ldots\nn \\
c(t,z) &=& 1 - \half z^4 \left(a_4(t) + \frac{3}{4} t a_4'(t)
\right) + \ldots \ .
\eea
The above expansion is expressed in terms of the single function $a_4(t)$,
which is not determined by the near-boundary analysis of the equations of
motion. This function is determined by the 
asymptotics of the full numerical solution. Holographic renormalisation
\cite{Balasubramanian:1999re,deHaro:2000xn,Skenderis:2002wp} is then invoked
to calculate the energy-momentum tensor 
(by construction this is of the form given in eq.~\rf{Tab}-\rf{Pr}). In this
way from the numerical solution at a given time $\tau$ one can read off the
energy density using the relation \cite{deHaro:2000xn}
\bel{holedens}
\epsilon(\tau) = -  \f{3 N_c^2}{8 \pi^2} a_4(\tau) \ .
\ee
It can be checked that the field equations imply that at early times the
expansion of $a_4(t)$ contains only even powers of $t$, which due to
\rf{holedens} implies an analogous property of the energy density  -- this was
established already in \cite{Beuf:2009cx} using Fefferman-Graham coordinates. 

Time evolution is computed using the 4th order Adams-Bashforth-Moulton
predictor-corrector approach. This turned out to be preferable to using
a Runge-Kutta integrator, due to the relatively large cost of spacial
integrations and solving eq.~\rf{linsys} at each time step. 

In practice setting up the numerical solution of Einstein equations is
somewhat subtle here, since one must choose a suitable range for the $z$
coordinate: ideally, it should cover the region starting at the boundary and
reaching just inside of the
event horizon. The location of the latter is however unknown until the entire
spacetime is determined. This chicken-and-egg problem is resolved in the spirit of
what is universally done in the numerical relativity literature: the apparent
horizon is located (if present), and the coordinate range is terminated
there. The position of the apparent horizon is computed directly from the
definition in terms of the expansion scalars (see for example
\cite{Booth:2009ct}). For our purposes the 
area of the apparent horizon is also of prime interest, since it is connected
to the entropy of the black brane (and ultimately, of the Yang-Mills theory
plasma). 

The position of the apparent horizon is not static, so one has to
dynamically adjust the range of $z$ every few time steps. Whenever this is done,
one has to translate the solution at that time to the new spacial grid. 
The heuristics for
this regridding are essential for the simulation to be stable long enough for
hydrodynamics to be reached. The accuracy of the numerical evolution can in
practice be monitored by checking whether the constraint equation (which is
satisfied at the initial time) remains
satisfied after each time step is taken.


\section{Initial states}
\label{sec.initial}

To specify a consistent initial condition for the asymptotically AdS geometry
one needs to satisfy the single 
constraint equation among \rf{einstein}. This equation involves only the
metric coefficients $b$ and $c$ 
\be
\half\left(\f{\p_zb}{b}\right)^2 - \f{\p^2_z b}{b} +
\half\left(\f{\p_zc}{c}\right)^2 - \f{\p^2_z c}{c} = 0 \ .
\ee
At $t=0$ we choose the function $b$ and then obtain the initial condition for
$c$ by numerically solving 
the above ordinary differential equation. Finally, the metric coefficient $a$
is obtained by solving the linear system \rf{linsys}.

The choice of $b$ has to be consistent with the near boundary expansion, whose
leading terms are given in eq.~\rf{expan}. In practice we chose various
families of functions (such as ratios of polynomials, elementary as well as
special functions) depending on some number of arbitrary coefficients which were then
randomly generated a number of times.

Note that the choice of $b$ at $t=0$ also determines the expansion of $a_4(t)$
in powers of $t$; this happens, because the near boundary expansion of $b$ is
expressed in terms of $a_4(t)$ and its derivatives. This series representation
of $a_4(t)$, valid close to $t=0$, can be used to control the numerical
solution at early times.

As discussed in the previous section, the initial energy density
when expanded in powers of the proper time around $\tau=0$ can only contain
even powers of $\tau$. It is not however clear whether one should assume that
the energy density is a nonzero constant at $\tau=0$. The color glass
condensate approach suggests that $\epsilon(0)=\epsilon_0>0$
(see for example \cite{Lappi:2006hq}), which  
was assumed in the present study. It could in
principle be that $\epsilon_0=0$, in which case the leading term would be 
\be
\epsilon(\tau)=\epsilon_{2k} \tau^{2k} + \dots 
\ee
with $\epsilon_{2k}\neq 0$ for some $k>0$. It has been suggested
\cite{Grumiller:2008va,Casalderrey-Solana:2013aba} that such
initial states (with $k=1$) appear in  
consequence of shock wave collisions in AdS. We hope to return to these kind
of initial states in the future, but in the work reported here we assumed that
the energy density is non-vanishing at the initial time. With this assumption one
can for convenience scale it to any given constant. We chose to fix
$a_4(0)=-1$. Using the relations \rf{efftemp} and \rf{holedens} this leads to
\bel{initemp}
\effti = \f{1}{\pi} \ .
\ee
While the initial geometries generated are very diverse, the states of
\sym\ which they map to all have the initial energy density scaled to the same
value. These initial states of the gauge theory are nevertheless distinct, as
their subsequent evolution shows.

\begin{figure}[!ht]
\begin{center}
\includegraphics[height = .25\textheight]{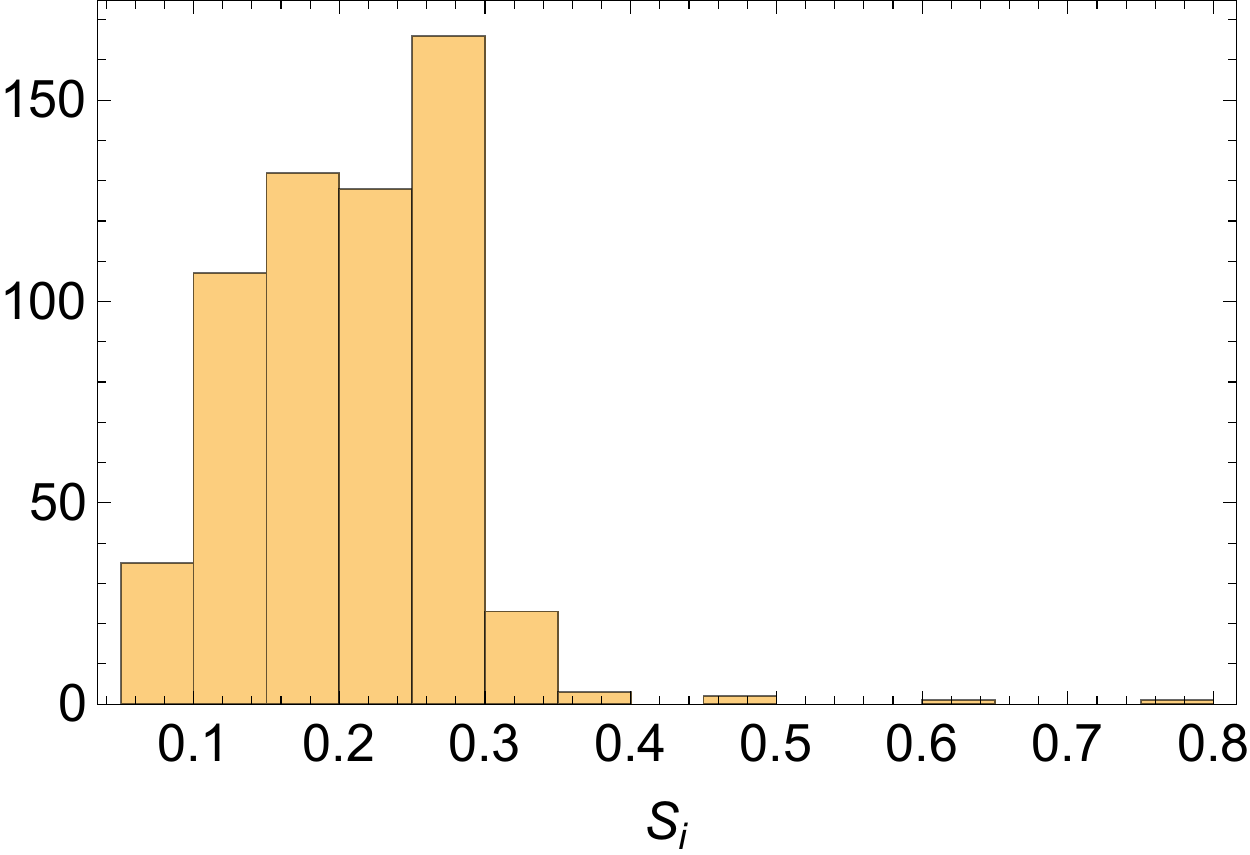}
\caption{The distribution of initial entropies.}
\label{fig.sinithist}
\end{center}
\end{figure}

A very interesting question is how these states should be characterized in the
Yang-Mills 
theory. Since these are highly non-equilibrium states one cannot use
thermodynamic notions to describe them. Indeed, it seems that one should not
expect any simple
description. However, as noted by Heller et al. \cite{Heller:2011ju}, from
a holographic perspective there 
is one characteristic which seems to be useful and interesting: it is
``initial entropy''.  This quantity can be defined unambiguously on the
gravity side in terms of the apparent horizon area of the black brane
geometry mapped to the boundary along null geodesics. This quantity is
guaranteed to be non-decreasing, and for large times becomes identical to
hydrodynamic (and ultimately thermodynamic) entropy. It is not a priori
obvious that an apparent horizon should always exist in the chosen time
slicing of AdS, but in all states considered here it exists either at the
initial time $\tau=0$ or very shortly thereafter. Using the
Bekenstein-Hawking relation to associate entropy with the apparent horizon
area one finds, with the assumed normalizations, 
\bel{entropy}
\bars = \f{a_{AH}}{\pi} \ ,
\ee
where $a_{AH}$ is the area of the apparent horizon. 
For all the numerical solutions considered this indeed coincides with the
thermodynamic result \rf{thydro} at late times. The range of entropies of all
the initial states generated is shown
in Fig.~\ref{fig.sinithist}.


\section{Thermalization}
\label{sec.thermal}

The results presented in this and the following sections follow from picking
initial gravity configurations as described above and evolving them until
times large enough for hydrodynamics to be valid.  

For each initial condition
the simulation gives the proper-time dependence of the effective temperature.
Similarly to the findings of
\cite{Heller:201e2j} we find that there are generically three different
behaviours. For states of initial entropy below $0.3$ 
typically the temperature either decreases monotonically from the start, or after an
initial plateau. For states with higher entropy the effective
temperature rises at first and only after some time the system begins to
cool. In \cite{Heller:201e2j} this phenomenon was referred 
as {\it reheating}. 

\begin{figure}[!ht]%
\begin{center}
\includegraphics[height = .19\textheight]{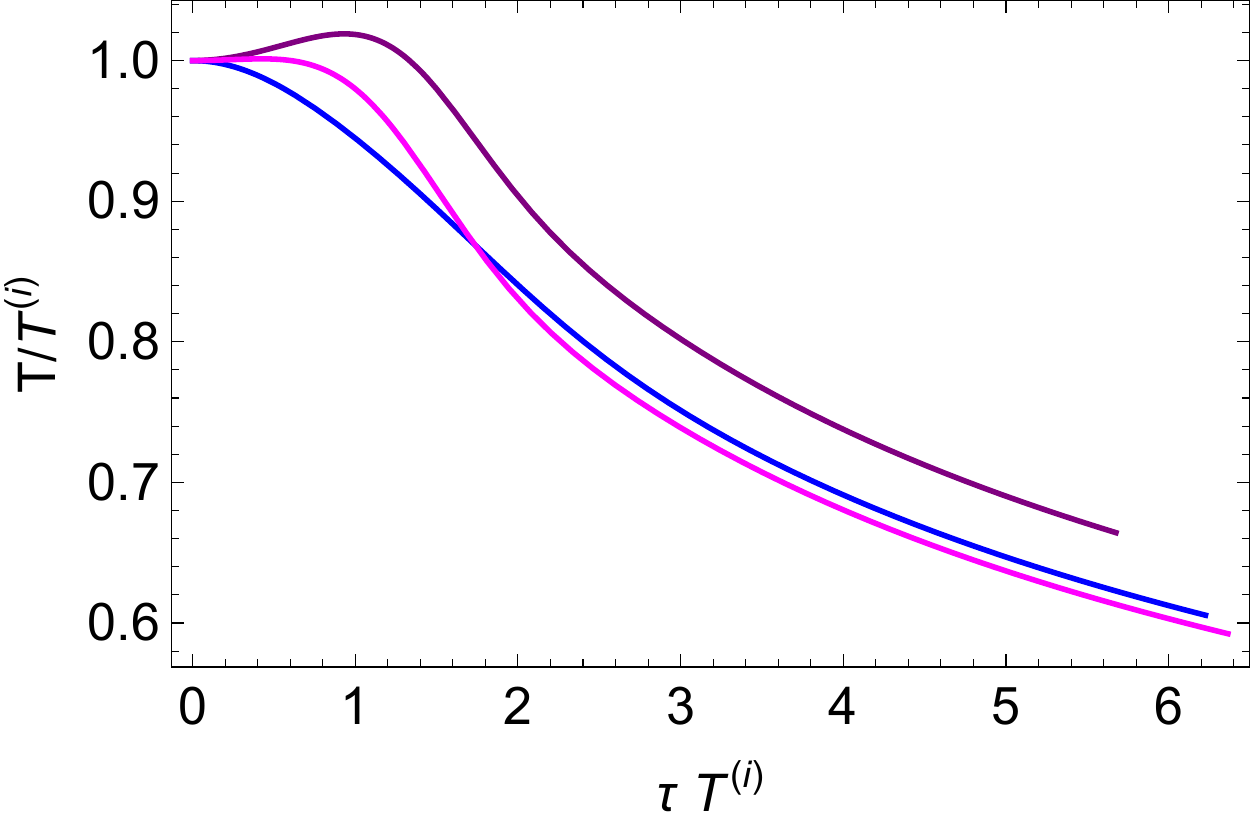}
\includegraphics[height = .19\textheight]{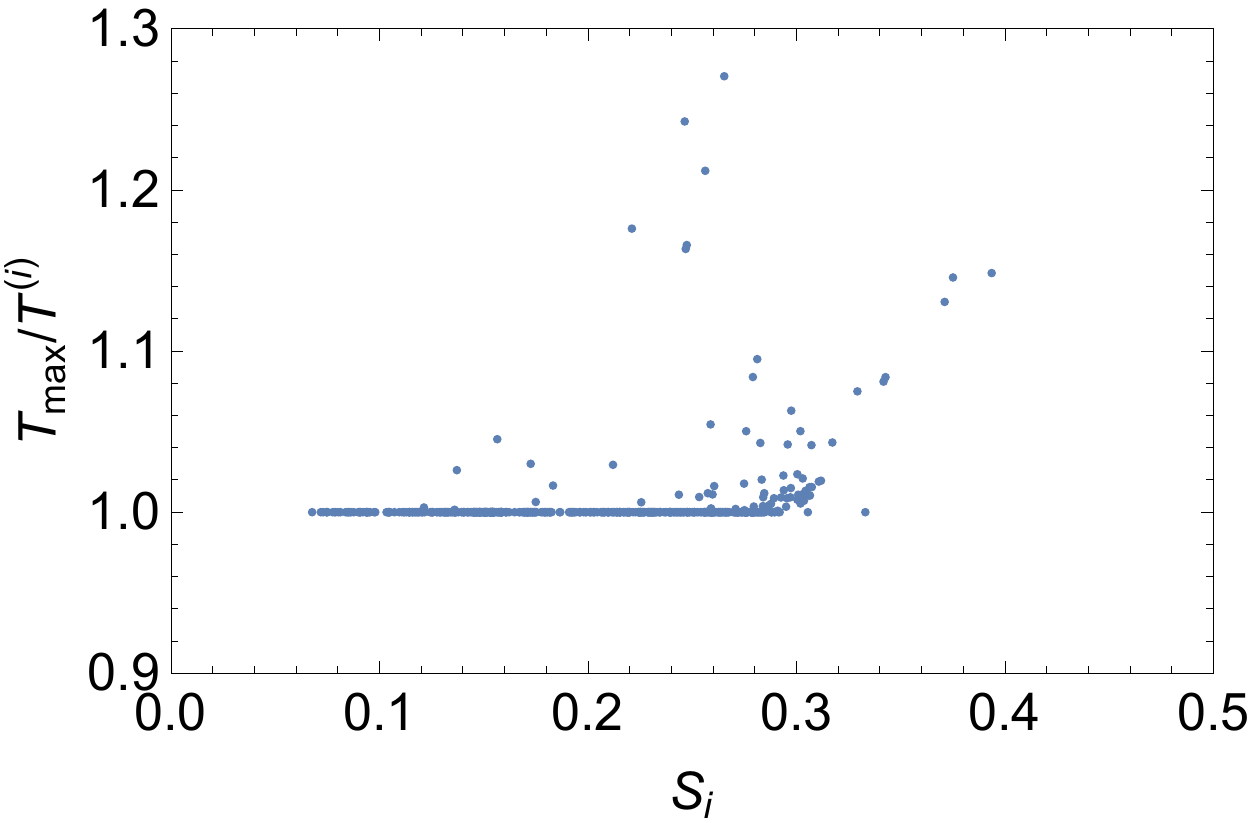}
\caption{Left panel: effective temperature as a function of time for three
  different initial entropies (both quantities in units of initial
  effective temperature). Right panel: maximal temperature (in units of
  initial effective temperature) as a function of the initial entropy.}
\label{fig.reheating}
\end{center}
\end{figure}

Since we have normalized the effective temperature to $1/\pi$ at $t=0$ for all
the initial configurations and for large times the system is cooling
in accordance with hydrodynamics, the maximal temperature is a measure of the
reheating phenomenon. Fig.~\ref{fig.reheating} shows that while reheating is absent
for almost all low entropy initial states, it is typical for entropies of
around $0.3$ or higher. There seems to be no
functional dependence of $T_{\rm max}$ on $\barsi$.


Despite these qualitative differences at early times, all numerical solutions
exhibit cooling at intermediate and late times. In the earlier studies of boost
invariant flow \cite{Heller:2011ju,Heller:201e2j} it was found that
hydrodynamics become a good 
description at some $w<1$ in all the 29 cases
considered. In the study presented here evolution was carried on until
$w=1.2$. In all 600 cases examined here hydrodynamics became an
accurate description significantly earlier, in agreement with
\cite{Heller:2011ju,Heller:201e2j}.  


\begin{figure}[!ht]
\begin{center}
\includegraphics[height = .25\textheight]{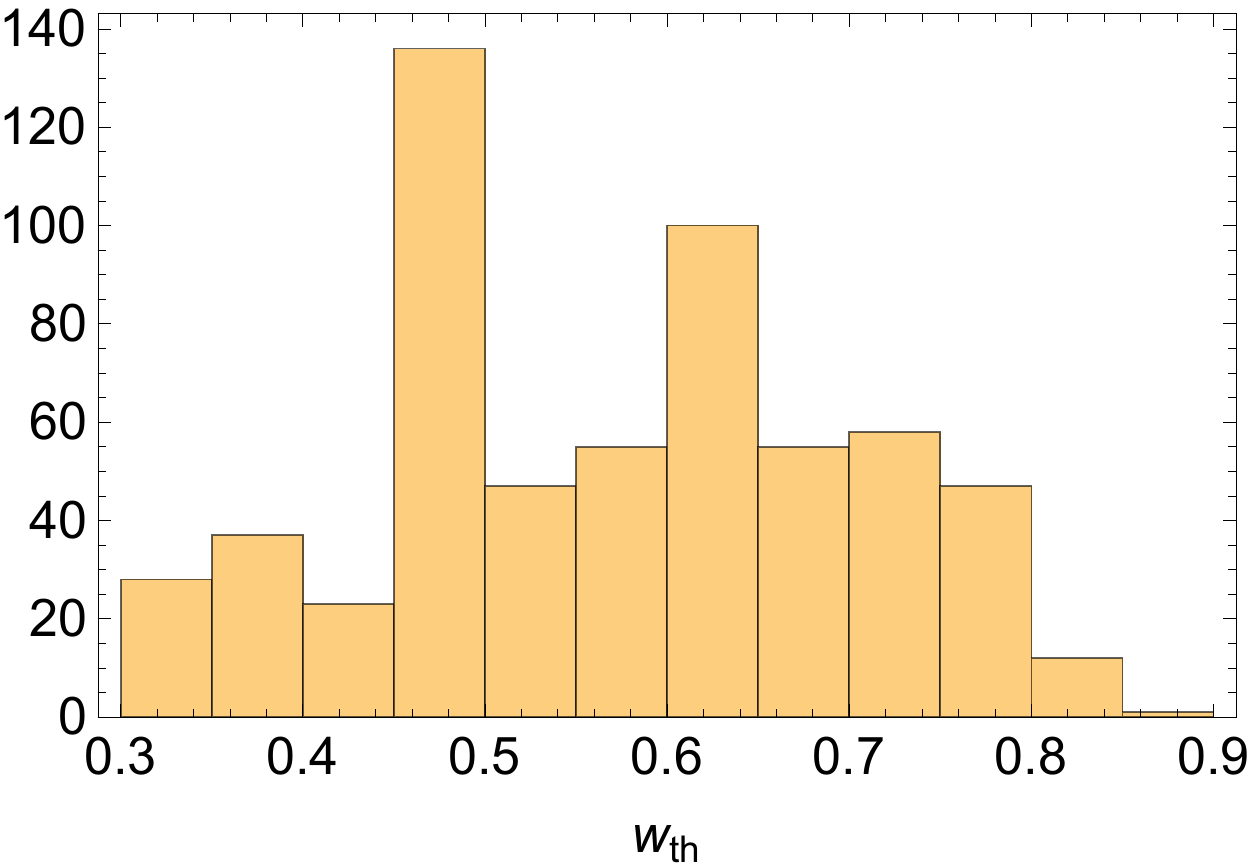}
\caption{The distribution of calculated (dimensionless) times $w=\tau T$ at
  the transition to hydrodynamics.}  
\label{fig.whist}
\end{center}
\end{figure}


The criterion for thermalization which we used is the same as that proposed in
\cite{Heller:2011ju}. Specifically, for a given solution $T(\tau)$ we evaluate the
corresponding function $f(w)$ defined by \rf{fdef}. We then calculate the
difference between this and the asymptotic form given by \rf{fhydro}. The
thermalization ``time'' $w_{th}$ is defined as the maximum value of $w$ such that
for $w>w_{th}$ 
\be
\Bigg|\frac{f(w)}{f_H(w)}-1\Bigg| < 0.005 \ .
\ee
The distribution of the values of $w_{th}$ found can be seen in the histogram
shown in Fig.~\ref{fig.whist}. It is clear from this that the early thermalization
in boost invariant flow observed in \cite{Heller:2011ju} is really a generic
feature 
of the dynamics.


\begin{figure}[!ht]
\begin{center}
\includegraphics[height = .25\textheight]{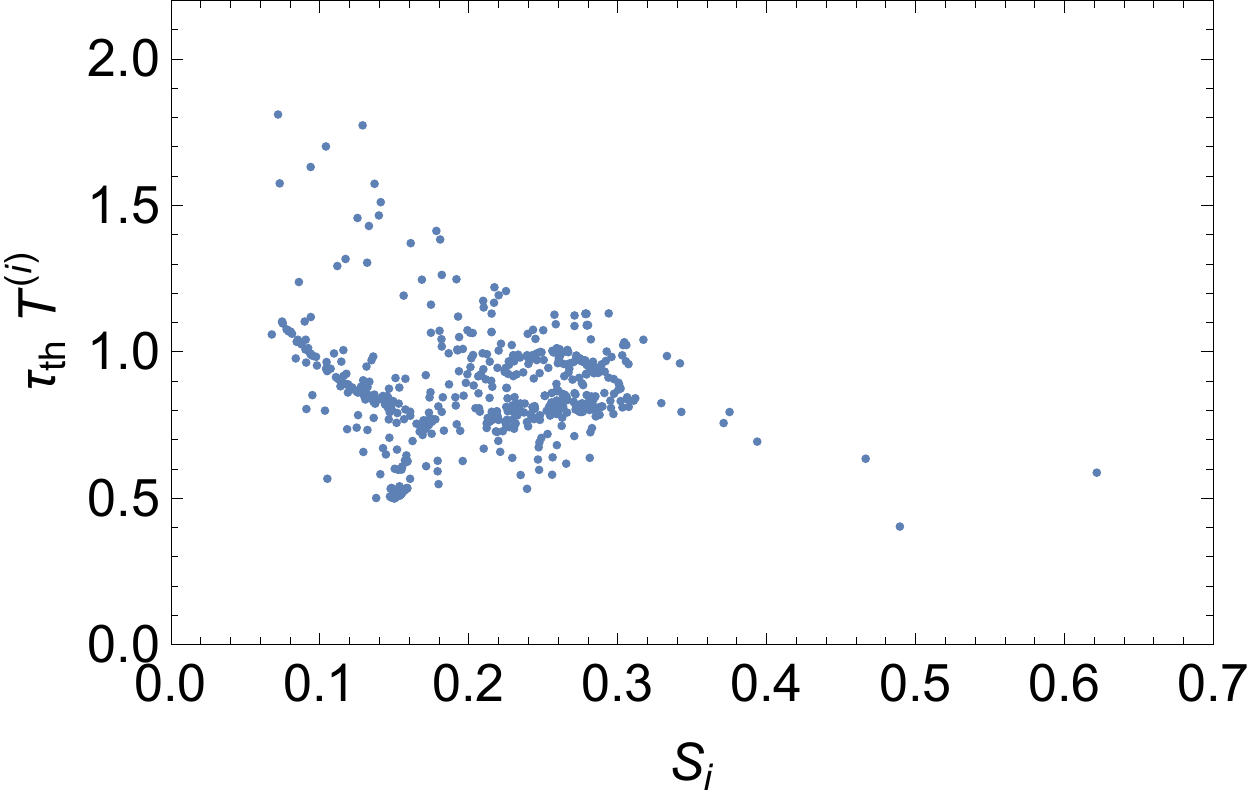}
\caption{Thermalization times in the units of initial effective
temperature as a function of the initial entropy.}
\label{fig.thermatime}
\end{center}
\end{figure}


The corresponding thermalization times are plotted against initial entropy in
Fig.~\ref{fig.thermatime}. The
units are those of initial temperature in order to easily compare with
reference 
\cite{Heller:2011ju}, where it was observed that for the sample of initial 
states considered in that work, there appeared to be a correlation between the
initial entropy and the thermalization time. In the larger set of solutions
examined here this does not appear to be 
the case.


\begin{figure}[!ht]%
\begin{center}
\includegraphics[height = .25\textheight]{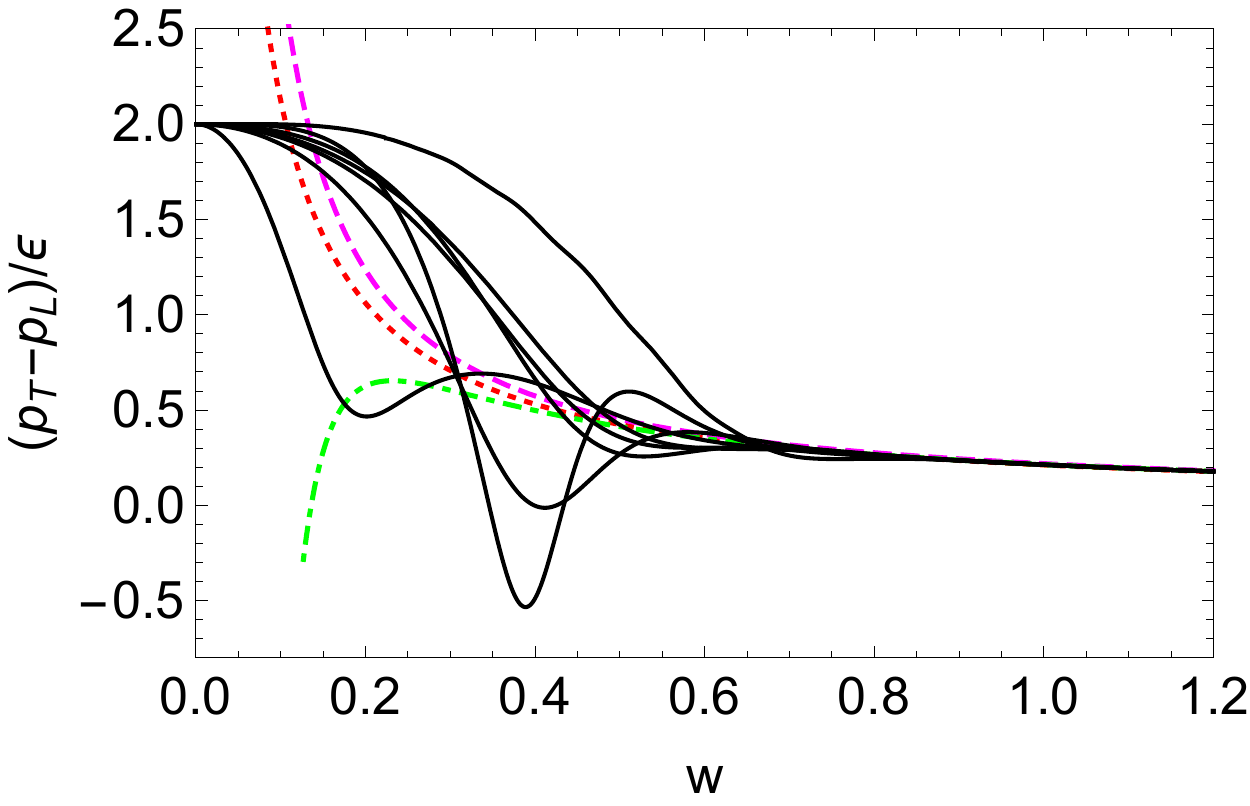}
\caption{The pressure anisotropy for a small sample of solutions (black
  curves). The remaining curves represent 1st (red, dotted), 2nd (magenta,
  dashed) and 3rd (green, dashed-dotted) order hydrodynamics.}
\label{fig.asym}
\end{center}
\end{figure}


At late times all the solutions converge to hydrodynamics, as expected. This
can be seen on plots showing the numerically computed function $f(w)$, or
equivalently by plotting the pressure anisotropy 
\be
\Delta \equiv \frac{p_T-p_L}{\epsilon} =  6 f(w) - 4 \ .
\ee
Typically this is quite large (on average $\Delta \sim 0.35$) at
thermalization, as seen in  
Fig.~\ref{fig.asym}. The conclusion is then that hydrodynamics becomes valid
rather early, and the large pressure anisotropy predicted by hydro at such
early times is in fact reliable. The oscillations visible on these plots as
hydrodynamics is approached indicate the presence of quasinormal modes
\cite{Heller:2014wfa}. 

\begin{figure}[ht!]%
\begin{center}
\includegraphics[height = .19\textheight]{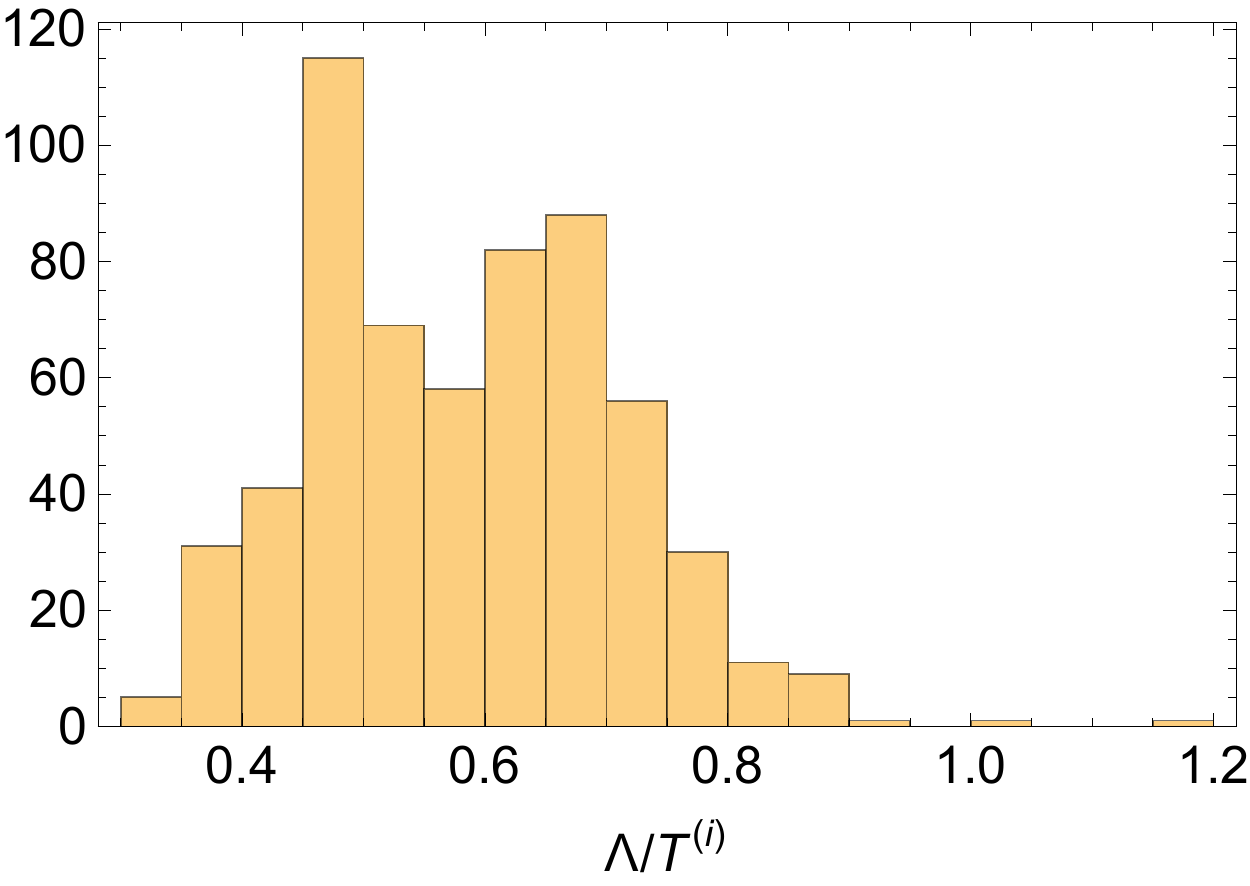}
\includegraphics[height = .19\textheight]{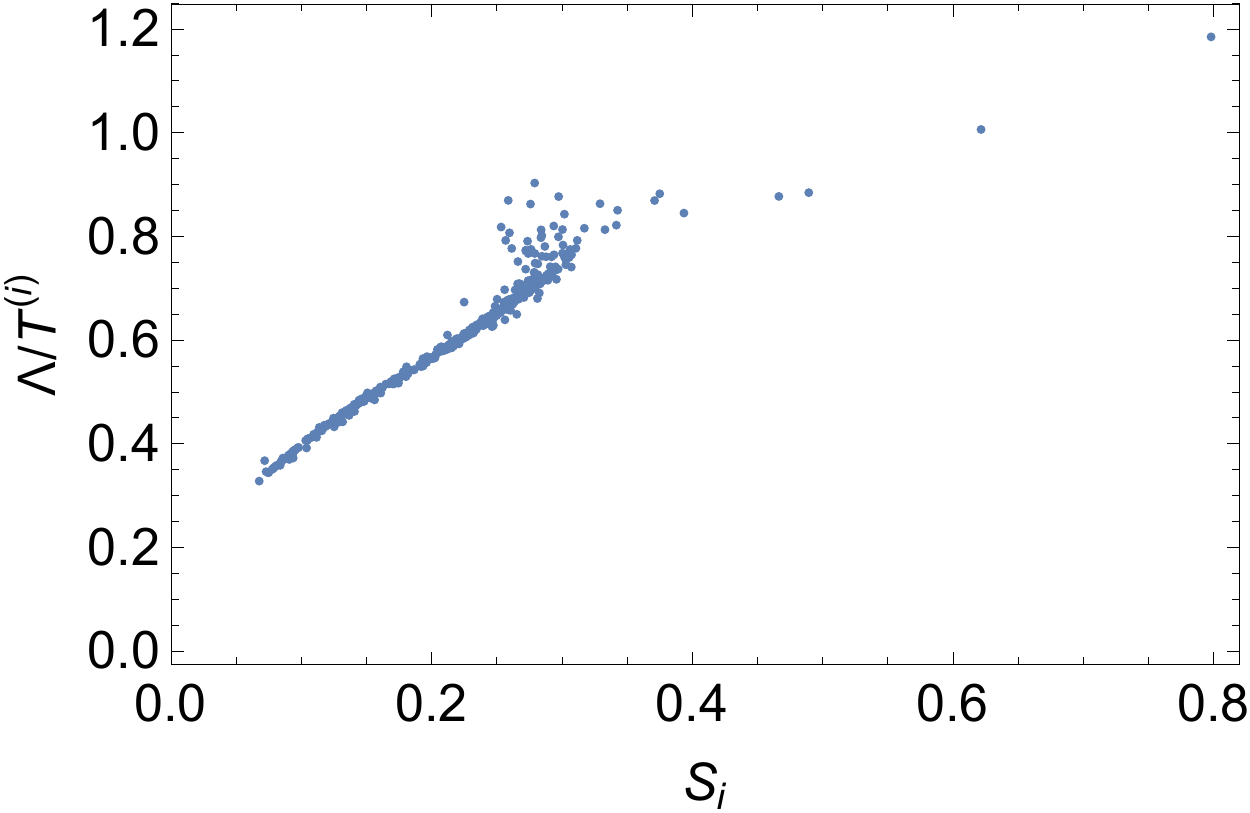}
\caption{Left panel: distribution of the fitted values of $\Lambda$. Right
  panel: fitted values of $\Lambda$ plotted against the initial entropy.} 
\label{fig.lambdas}%
\end{center}
\end{figure}

After thermalization is attained the hydrodynamic description becomes valid
(by definition). As discussed in section~\ref{sec.bif}, the boost-invariant
hydrodynamics of a conformal fluid is determined by the single scale
$\Lambda$. For any numerical solution this value can be found by fitting
formula \rf{thydro} to the tail of the numerical data. A histogram of the
values of $\Lambda$ obtained in this way is shown in
Fig.~\ref{fig.lambdas}. 

The distribution seen in the left panel of Fig.~\ref{fig.lambdas} is
reminiscent of the histogram of initial entropies. In fact, the simulations
reported here suggest that the scale $\Lambda$ is linearly correlated with the
initial entropy. This is seen in the right panel of Fig.~\ref{fig.lambdas}.
Since in QCD the final charged particle multiplicity is proportional to the
final entropy (which is $\Lambda^2$ here), this result suggests that this
multiplicity is related to the {\em square} of the initial entropy\footnote{Of
  course, unless one understands the meaning of initial entropy directly in
  field theory terms, this observation is not very useful.}.


One can see that the linear correlation between $\Lambda$ and the initial
entropy is valid for a large range of initial
entropies, but at the high end the correlation seems to be lost. This suggests
that in this regime the initial entropy is not the dominant
characteristic of the initial state, and other factors come into
play.


\begin{figure}[!ht]
\begin{center}
\includegraphics[height = .25\textheight]{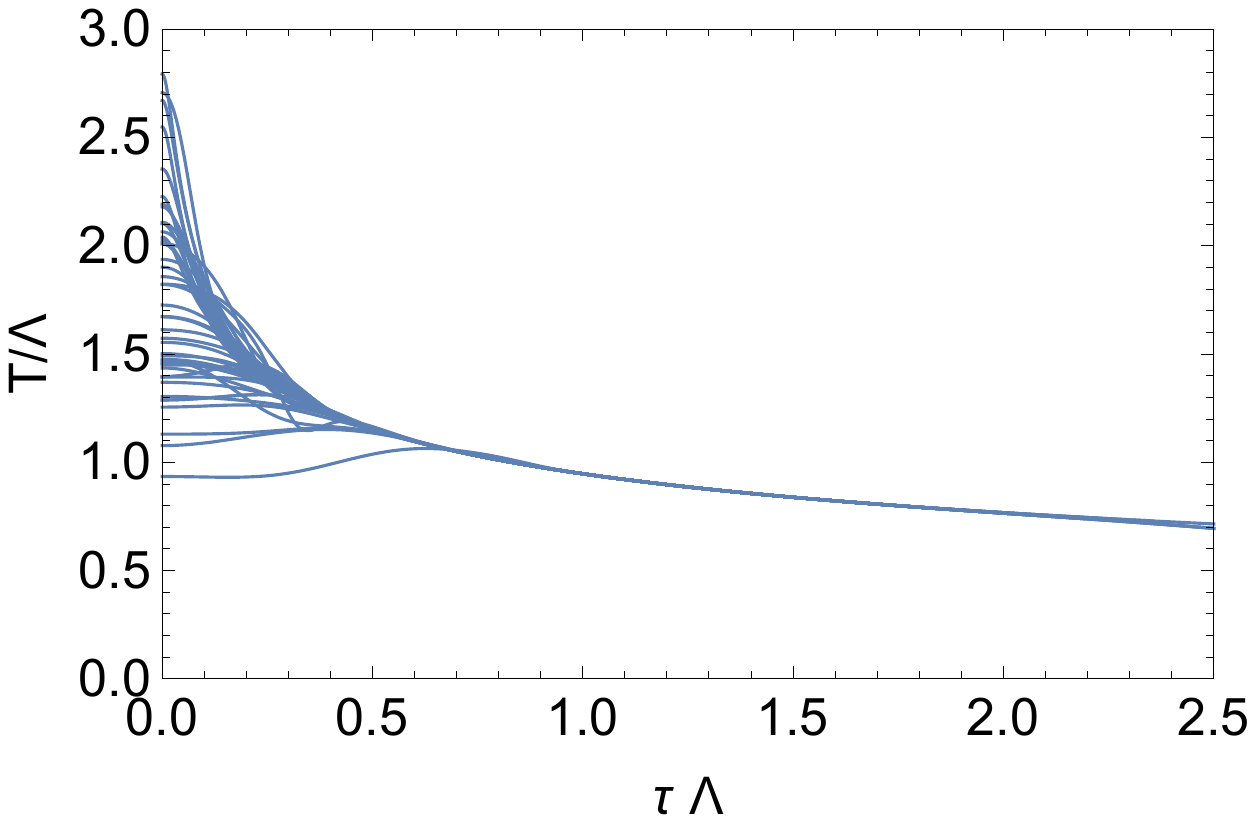}
\caption{Time dependence of the rescaled effective temperature.}
\label{fig.alltemp}
\end{center}
\end{figure}


Finally, it is instructive to view the hydrodynamization process from
a slightly different perspective \cite{Janik:2014zla}. Having computed the scale 
$\Lambda$, one can rescale the proper time and effective
temperature in such a way that the rescaled data reach hydrodynamics with
$\Lambda=1$. Such a rescaling amounts to normalizing the data so all solutions
have the same final entropy.
If this rescaling is done for all solutions they will start out
at various values of the initial effective temperature and will all
converge at late times. In Figure \ref{fig.alltemp} we see this for a few
sample configurations.


\section{Entropy}
\label{sec.entropy}

The observed viscosity of quark-gluon plasma implies that entropy rises during
hydrodynamic evolution, as well as in the pre-hydrodynamic stage -- to the
extent that it makes sense to speak of entropy in a highly non-equilibrium
system. In the case when a holographic dual is available we do have a notion of
entropy as defined by the apparent horizon area, so we can quantify entropy
production at all times (at least in those cases where an apparent horizon
exists and was identified within the computational grid).

\begin{figure}[!ht]%
\begin{center}
\includegraphics[height = .25\textheight]{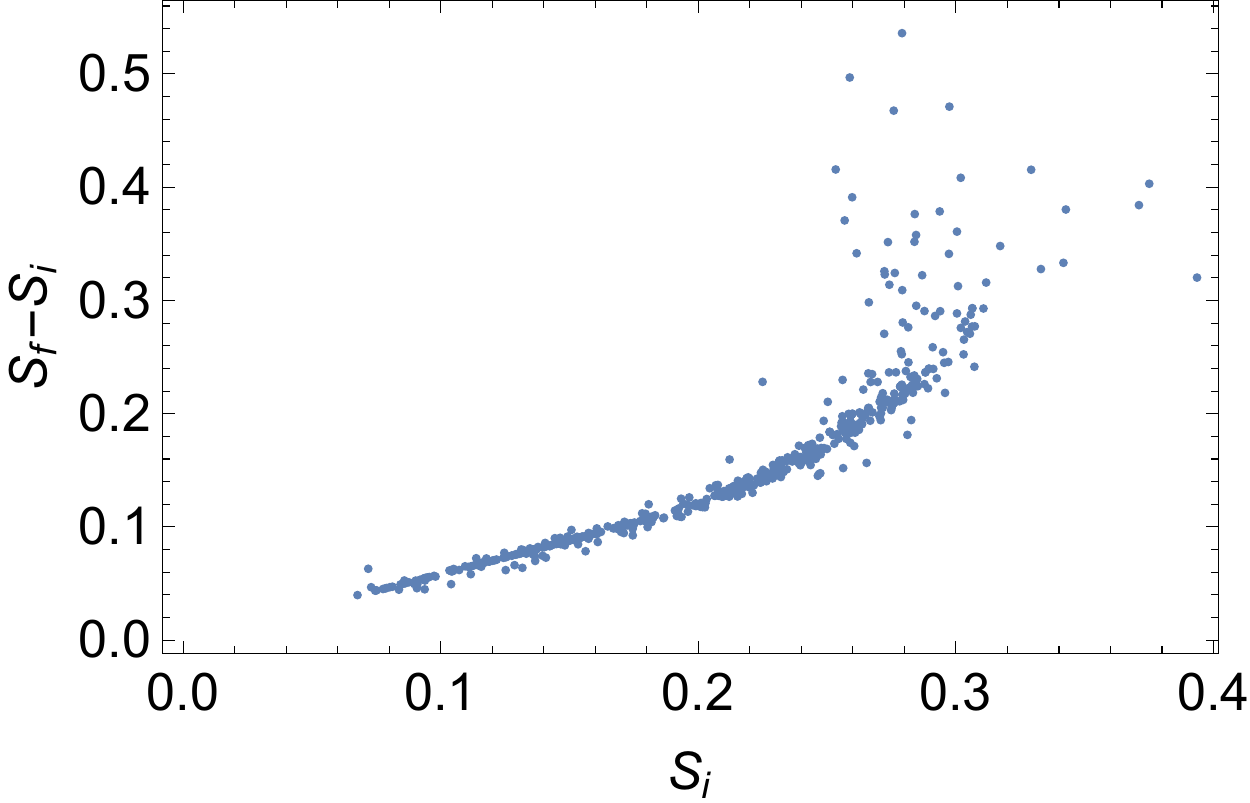}
\caption{
Entropy production as a function of initial entropy.
}%
\label{fig.eprod}%
\end{center}
\end{figure}

Entropy production is an important aspect of the dynamics -- it 
is related to particle production in heavy ion collisions.
An interesting observation was made in reference \cite{Heller:2011ju},
whose authors pointed out that the entropy produced is correlated with the
initial entropy. In fact, for Bjorken flow the final entropy 
is given in terms of the scale $\Lambda$ by eq.~\rf{ethermo}, so using
eq.~\rf{initemp} one has
\bel{entfina}
\barsf = \pi^2 \Lambda^2.
\end{equation}
As explained earlier, the actual value of $\Lambda$ is obtained by fitting the
3rd order 
formula \rf{thydro} to the tail of the numerical solution. 
This relation, together with our previous observation of a linear correlation
between $\Lambda$ and the initial entropy $\barsi$, implies that for
small initial entropies there should be a correlation between
the entropy produced, $\barsf - \barsi$, and the initial entropy
$\barsi$, which can be fitted by a quadratic function. 
This is indeed seen in the data obtained in our simulations (see
Fig.~\ref{fig.eprod}) for evolution from initial states with entropy 
below $0.3$. For initial states with higher entropy  the correlation is lost. This
is of course consistent with our earlier observations concerning the
appearance of a ``chaotic phase'' when the initial entropy exceeds $0.3$.


It is also interesting to consider how much entropy is produced during
the hydrodynamic stage of evolution. Since hydrodynamics provides an accurate
description of the dynamics already at rather early times, it is to be
expected that a significant part of the final entropy is due to hydrodynamic
evolution -- more than often assumed. This can be taken as motivation for
efforts to resum the hydrodynamic series
\cite{Lublinsky:2007mm,Lublinsky:2011cw,Heller:2013fn,Bu:2014sia,Bu:2014ena}.  

In the following, by entropy at thermalization (denoted by $S_{th}$) we mean
the apparent 
horizon entropy given by eq.~\rf{entropy} evaluated at thermalization time.
It is interesting to estimate the size of gradient corrections to
thermodynamic entropy at this stage of evolution. This can be done by comparing
the ``exact'' entropy given in terms of the apparent horizon area by
eq.~\rf{entropy} with hydrodynamic entropy defined by the entropy current
within the gradient expansion \cite{Booth:2009ct}. Specifically, using
eq.~\rf{eqed} and \rf{entperura} one finds that the leading order
(thermodynamic) approximation to the entropy defined in eq.~\rf{entrat} is
\bel{entpf}
S^{(0)} = \tau \pi^2 T^3 \ , 
\ee
assuming the normalization \rf{initemp}. Gradient
corrections to hydrodynamic entropy are then quantified by the ratio
$(S-S^{(0)})/S$. In the simulations described here we find that at
thermalization time this ratio assumes values in the range 0.05 -- 0.15. This is
consistent with the fact that corrections to the leading order term,
eq.~\rf{entpf}, first appear at 2nd order in the gradient expansion
\cite{Booth:2009ct}. For this reason, 
even though the pressure anisotropy is large at thermalization time, the
gradient corrections to thermodynamic entropy are relatively
small. Therefore, in the situation studied here, thermodynamic entropy
is a reasonable approximation as soon as 
hydrodynamics becomes valid. 


\begin{figure}[!ht]%
\begin{center}
\includegraphics[height = .19\textheight]{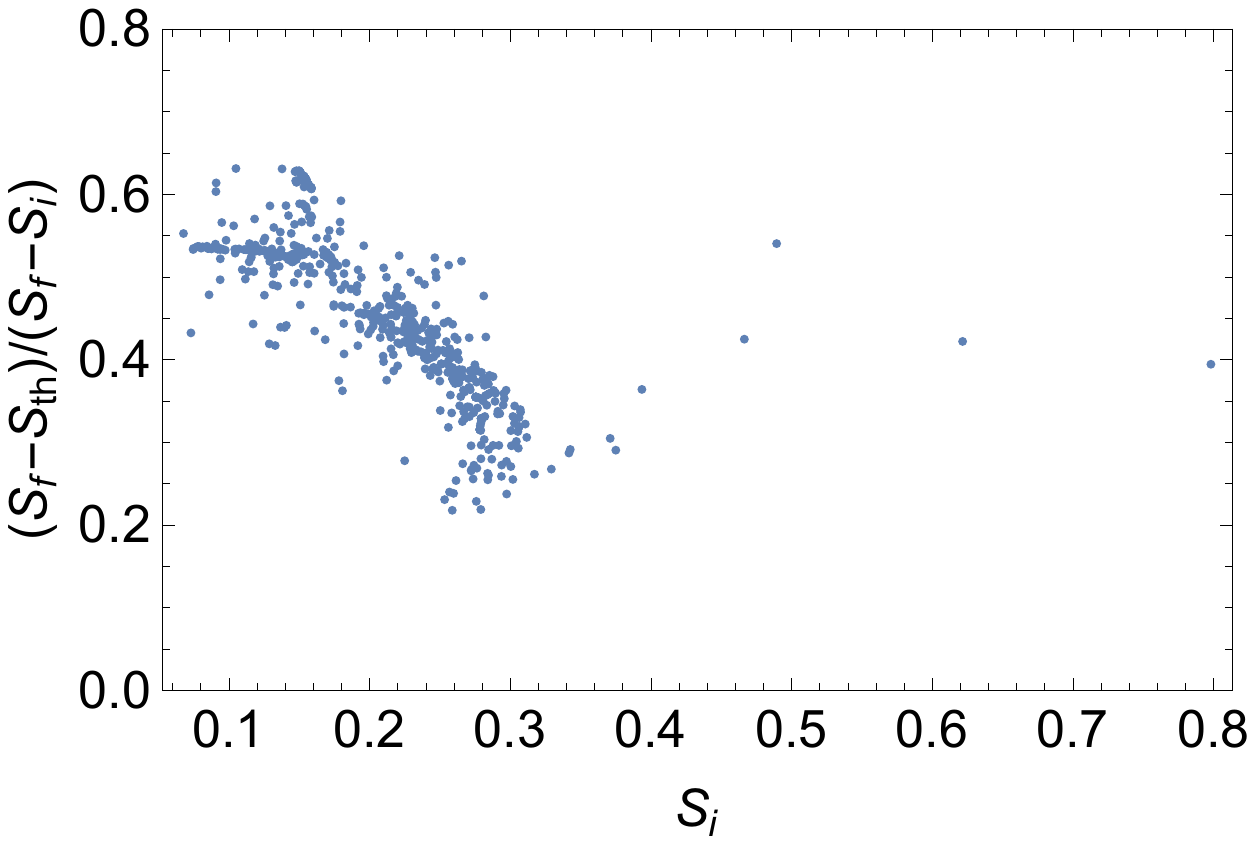}
\includegraphics[height = .19\textheight]{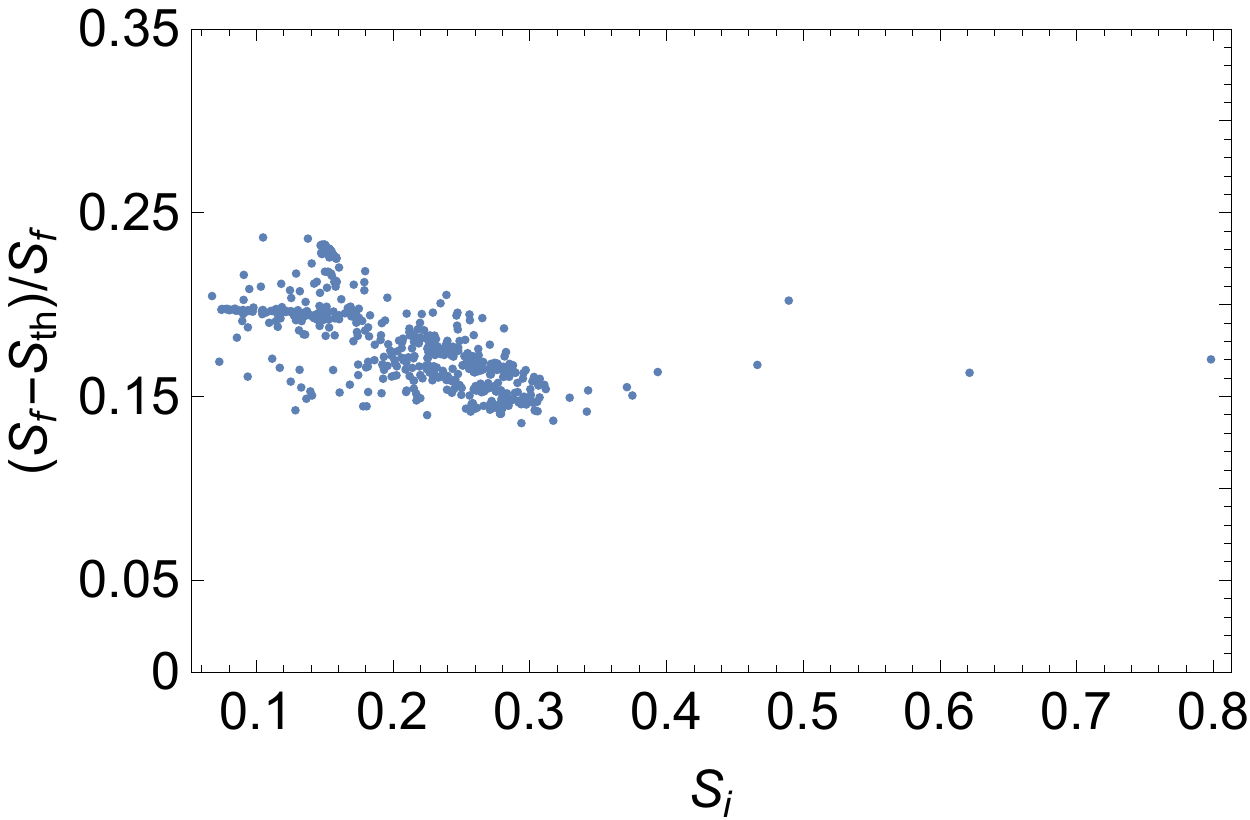}
\caption{
Entropy production during the hydrodynamic expansion, shown as ratio relative to total
entropy produced during evolution (left panel) and to final
entropy (both as functions of initial entropy).
}%
\label{fig.eprodd}%
\end{center}
\end{figure}


Entropy production at different stages of evolution is illustrated in
Fig.~\ref{fig.eprodd}. The left panel shows the ratio of entropy produced
during the hydrodynamic stage to the total entropy production. This can be
very high, especially for states of low initial entropy, for which often more
than half of the entropy produced comes from the hydrodynamic
evolution. Usually however, one compares the entropy produced during the
hydrodynamic stage not to the entropy produced during complete evolution, but
rather to total entropy. This ratio is plotted in the right panel of
Fig.~\ref{fig.eprodd}. This quantity is more in line with expectations,
although it can still be quite significant: about $15\%-25\%$ of the total
entropy is produced by the dissipative effects during hydrodynamic evolution.



\section{Conclusions}
\label{sec.conclusion}

We have analyzed the time evolution of 600 randomly generated far from
equilibrium initial states of strongly coupled \sym\ plasma well into the
hydrodynamic regime.  It is important that in the approach presented here the
number of initial states is not limited by anything else than patience, and
more solutions could easily be generated.  

Having a fairly large sample makes
it possible to look for generic features of the dynamics. While the selection
of initial states does not have any built-in bias that we are aware of, we
cannot claim that the set we used is actually representative. One possible
limitation is that some randomly generated initial geometries do not evolve in
a stable way for long enough to reach the hydrodynamic regime. This is most
likely because the heuristics we use to estimate the position of the event
horizon are not always effective, especially in cases where no apparent
horizon exists on the initial time slice. This could be the reason why among
the initial states found in \cite{Heller:2011ju} there are some with lower initial entropy
than what we have found.

The main motivation for this study was to see how generic is the early
thermalization observed in the limited sample of initial states considered in
\cite{Heller:201e2j}. Our results indicate that early thermalization is the
norm. Hydrodynamic behaviour typically sets in at a time of the order of 0.6
over the effective temperature, when the pressure anisotropy is still
significant (with $\Delta$ of the order of 0.3).  At the onset of
thermodynamics one can estimate the entropy in field theory using the leading
order hydrodynamic approximation, which we have verified to be reasonably
accurate, as expected \cite{Booth:2009ct}. This serves as a reference point to
quantify the amount of entropy produced during the hydrodynamic stage of
evolution -- the result is that the increase in entropy is typically about
20\% of the final entropy, which amounts to about 40\% of the total entropy
produced during evolution.

In the previous study by Heller et al. \cite{Heller:2011ju} it was pointed out
that in the context of holography one could try to characterize far from
equilibrium initial states
by their entropy measured by the area of the apparent horizon appearing in the
dual geometry. Our numerical experiments confirm that for a 
range of initial entropies this is indeed a useful concept. One of the
surprises that have emerged was that the parameter $\Lambda$ which sets the scale
for the hydrodynamic tail is linearly correlated with this initial
entropy. However, for large entropies (larger than $0.3$ in our units) this
relation is lost and some ``chaotic'' behaviour develops. This indicates that
for such states there are other characteristics which are important.  Even for
initial entropies which are not so high, it is not true that they 
fully characterize the initial state, because thermalization times are not
correlated with them at all. 

The fact that initial entropy seems to be a useful concept raises the question
of its field theory meaning. In other words, how can one define it without
appealing to the holographic picture? Perhaps this could 
be addressed in the context of models of the initial
post-collision state. Interesting attempts going in this
directions have already been made in \cite{Peschanski:2012cw,Kutak:2011rb}. 
The connection between the notion of dynamical entropy defined in these references
and apparent horizon entropy is not clear, but could perhaps be
made via an approach such as \cite{Iancu:2014ava}. 

Apart from pursuing answers to the questions discussed above, the work
reported here could be 
generalized in various ways. One direction would be to consider
an alternative class of initial conditions, with different early time
behaviour. Another natural avenue to pursue is the behaviour of
non-local observables (such as Wilson lines or entanglement entropy
\cite{Pedraza:2014moa}) in the expanding backgrounds described here. We hope
to return to these issues in the future.

\vspace{15pt}
\noindent {\bf Acknowledgments:} The authors would like to thank David
Blaschke, Micha\l\ P. Heller, Romuald Janik, Daniel Nowakowski, Przemys\l aw
Witaszczyk and Wilke van der Schee for valuable discussions. This work was
partially supported by Polish National Science Center research grant
2012/07/B/ST2/03794 (GP and MS) and a post-doctoral 
internship grant No. DEC-2013/08/S/ST2/00547 (JJ).

\bibliography{b}{}
\bibliographystyle{utphys}

\end{document}